# Majorana Bound State Cavities

Babak Bahari,[1] Jae-Hyuck Choi,[1] Yuzhou G. N. Liu,[1] and Mercedeh Khajavikhan[1*]

[1] *Ming Hsieh Department of Electrical and Computer Engineering, University of Southern California, Los Angeles, California 90089, USA*

e-mail address: khajavik@usc.edu

Cavities play a fundamental role in optical physics by providing spatio-temporal confinement of energy that facilitates light-matter interactions. They are routinely utilized in a variety of settings, ranging from lasers to spectral filters, to nonlinear wave mixers, and modulators. While their resonant properties make them suitable for sensing, in many applications, like in lasers, it is desired to alleviate their sensitivity in order to make a device more resilient to perturbations. Along these lines there have been several recent reports of using concepts from topological physics in order to design laser arrays that are inherently more robust. Here, we present a new class of topological cavities based on Majorana bound states that provide unique scaling features as well as non-degenerate single-mode behaviors. These cavities may lead to a new family of lasers and laser arrays that are robust to defects, fabrication imperfections, and external perturbations.

## I. Introduction

In 1937, Ettore Majorana speculated a new type of fermionic particles that are their own antiparticles, now widely known as Majorana fermions. In condensed-matter physics, Majorana fermions can appear as quasiparticle excitations that are governed by non-abelian statistics [1-3]. Since they tend to emerge at the center of the bandgap as solutions with zero eigen-energy, they are also referred to as Majorana zero modes or Majorana bound states (MBS). In the past few decades, several experiments have provided evidence as to solid-state manifestations of such bound quasi-particles in 1D semiconductor-superconductor hybrid systems [4-9]. These zero modes were also found as localized states at both ends of InSb semiconductor nanowires and iron nanochains [10-12]. Because of their unique properties, especially their resilience to noise, in recent years Majorana bound states have received considerable attention for applications in the growing areas of quantum computing and quantum error-correcting codes [13-15].

While MBS has been extensively studied in topological superconductors and other condensed matter systems, they have only been considered recently in bosonic arrangements [16-20]. In 2016, Iadecola et al. proposed a realization of 2D topological Majorana bound states in a photonic setting based on noninteracting photons propagating in a nontrivial background with topological defects whose position are controllable [16]. More recently, a few experiments verified the presence of such modes in photonic arrangements using arrays of coupled waveguides. In these systems, light can channel into Majorana topological guided modes and braid with the accumulation of non-Abelian phases [17,19]. Photonic Majorana bound states have also been demonstrated in integrated platforms using silicon nanostructures [20]. These modes are fundamentally different from other zero modes appearing in arrays with Su-Schrieffer–Heeger (SSH) model [21-24] or other higher order topological systems since they require analogous semiconductor-superconductor type of interactions.

Here, we introduce a new class of topological photonic cavities based on Majorana bound states by exploring the prospect of utilizing vortex distortion in a hexagonal lattice-also known as Kekulé texture [25]. Due to the topological nature of MBS, these cavities are expected to exhibit some degree of robustness to structural errors, thus may find applications in laser physics [26-32]. Unlike the standard topological insulator lasers in which the non-trivial mode appears at the boundary of the lattice, Majorana bound states reside at the center of the structure. Moreover, by nature, these modes are non-degenerate, appearing as single states at zero-energy. In our design based on high refractive index III-V semiconductor cylinders, the MBS mode is strongly confined with a large mode/gain overlap and shows a large quality factor when compared to any other mode inside the bulk of the photonic lattice. This allows to implement a new type of single mode broad area laser. This is also in sharp contrast with previous design in [20], where the mode mainly occupies the air region and minimally overlaps with the semiconductor gain material, thus cannot reach the threshold of lasing. In addition, for the first time, we systematically study the robustness of MBS modes using perturbation theory. We demonstrate that MBS is quite resilient when the coupling strength between the constituent cells of the array is randomly varied, while all the other bulk modes experience large fluctuations in their eigen-energies. The MBS cavity also appears to be quite scalable, thus allowing for extraction of larger output power when used as a laser array. The Majorana bound state topological cavities can be used in applications where less sensitivity and pure single-mode operation is desired, in order to maintain their functionality despite fabrication errors or external perturbations.

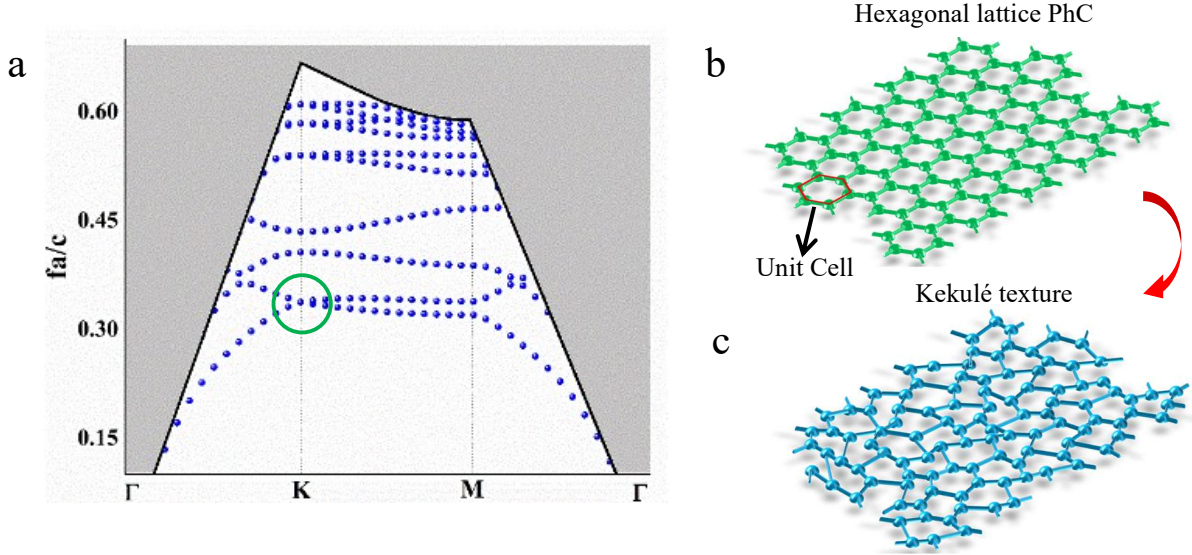

FIG. 1. a) Band Diagram of a PhC with hexagonal unit cell. The structure comprises of InGaAsP cylinders that are connected to each other by small bridges. The periodicity and the radius of these cylinders are $p$ and $0.2p$, respectively. A Dirac Cone is formed at K point (inside the green circle). b,c) The hexagonal lattice PhC before (b) and after (c) applying Kekulé distribution.

## II. Photonic Majorana Bound State Cavities

It is known that graphene-type lattices exhibit accidental crossings and form Dirac cones in their reciprocal space. Figure 1a shows the band diagram of a photonic crystal (PhC) unit cell made of InGaAsP cylinders in a hexagonal arrangement. These cylinders are connected to each other using short bridges as shown in Fig. 1b. The periodicity, radius of cylinders, and width of bridges are $p = 500$ nm, $r = 0.2p$, and $w = 0.1p$, respectively. The thickness of cylinders was taken to be $200$ nm. In the band diagram shown in Fig. 1a, the first two lowest out-of-plane modes touch each other at the K point (highly symmetric point) and form a Dirac cone at a wavelength of $\lambda = 1550$ nm.

Generally, any effect that breaks the inversion and/or time reversal symmetry of a hexagonal lattice can result in a bandgap opening. In optics, it is also possible to open a bandgap without breaking these symmetries [25, 33]. In 2007, Hou et al. [25] proposed a vortex distortion to this lattice and deformed it based on a Kekulé transformation, thus opening a spectral gap in the massless Dirac point spectrum of the graphene structure. The resulting Kekulé texture happens to be mathematically analogous to a superconducting order parameter. This in turn leads to the formation of Majorana zero modes that display topologically non-trivial characteristics (Fig. 1c) [34]. Using this approach, a fractionalized charge can be generated in the 2D lattice that has a zero energy while the time-reversal symmetry remains invariant.

Within the tight-binding formalism, the Kekulé distortion can be described by the following Hamiltonian:

$$H = -\sum_{r \in A} \sum_{i=1}^{3} (\kappa + \delta\kappa_{r,i}) a_r^\dagger b_{r+i} + H.c., \quad (1)$$

where $\kappa$ is the coupling between each two elements and $\delta\kappa$ is the coupling detuning due to Kekulé transformation. Here,

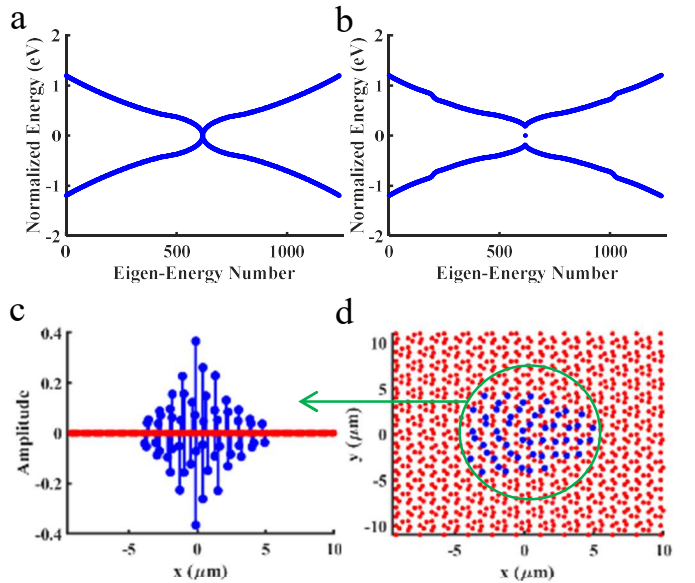

FIG. 2. a,b) The eigen-energy of all modes of an array with hexagonal arrangement (a) and Kekulé texture (b). A bandgap is opened and Majorana bound state with a zero-energy is at the center of bandgap. c) Amplitude of the field profile of a MBS mode. d) Top view of the field profile. Red color shows the position of cells, and blue color shows cells with nonzero amplitude.

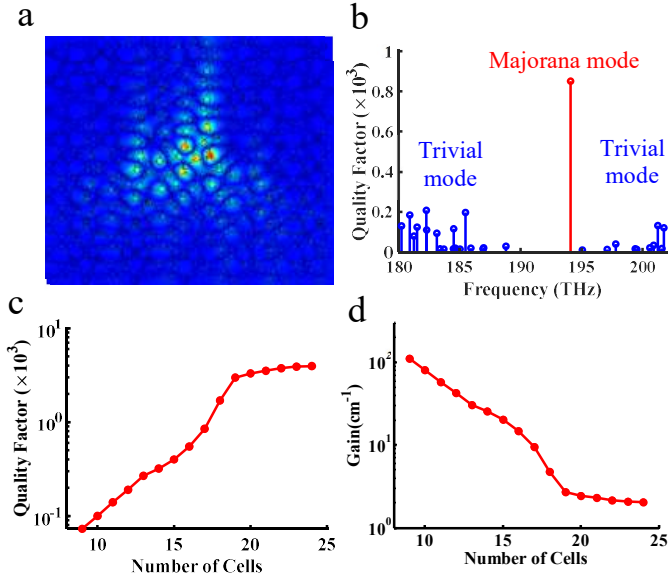

FIG. 3. a) Electric field intensity profile of the structure with Kekulé texture using full wave simulation. b) Eigenmodes of the structure. MBS mode has the highest-Q mode (~900) over a broad frequency range compared to trivial modes (<200). c) Evolution of the quality factor of MBS mode as a function of the array size. d) Evolution of the minimum gain required for laser as a function of the array size.

The eigen-energy of the modes of this structure are numerically calculated using the Hamiltonian of the Kekulé texture written for a finite size structure. Figure 2a shows the band diagram of the PhC with a hexagonal arrangement (where no Kekulé transformation is applied). Clearly, a Dirac point (level crossing) is observed at the zero-energy level. However, after applying the Kekulé effect, while the time-reversal symmetry is preserved, the coupling strength between each site and its nearest neighbors is modified in such a way that a vortex is generated. Figure 2b shows the band structure of the Kekulé deformed lattice, where a zero-energy mode appears at the middle of the opened bandgap. This mode represents a Majorana state that is bounded to the zero energy. The field intensity distribution associated with this mode is shown in Fig. 2c and 2d. The mode is confined at the center of the vortex (structure) and its intensity decays as one moves away from this point. Here we used the vorticity of 1 (coefficient of $\phi$), but any positive or negative integer can be used. A vorticity with an absolute value larger than unity can generate multiple zero modes inside the bandgap.

The size of the Majorana bound state mode is controllable by adjusting parameter $d(r)$, as it controls the opening of the bandgap. We should note that the zero mode fully resides in one of the sublattices (either $A$ or $B$, dictated by the sign of vorticity). This is evident from the full wave simulation using COMSOL Multiphysics software. When solving Maxwell's equations, a defect mode is found that is confined at the center of the vortex (Fig. 3a). For this structure, other trivial eigenmodes are also calculated (shown in Figure 3b). Over a broad frequency range, a Majorana bound state remains as the most confined mode exhibiting the highest quality factor (defined as $Q = Re\{\tilde{f}\}/2Im\{\tilde{f}\}$ where $\tilde{f}$ is the complex eigen-frequency), and it can be scaled up as a function of the array size as depicted in Fig. 3c. This is because the MBS mode remains very localized at the center of Kekulé lattice. The size of vortex is independent of the array size as far as the $d_0$ and $\zeta$ parameters stay unchanged. Even after increasing the size of the structure, the cavity remains single-moded. The physical extension of the lattice only affects the position at which the vortex is terminated, thus the tail of the mode can touch the boundary and change the quality factor. However, this only marginally affects the zero-mode due to its topological properties. On the other hand, the trivial modes (that generally show significantly lower quality factors) are distributed over the entire array, and their quality factors tend to dramatically fluctuate with the size of the structure. In Fig. 3c, there is a sharper transition of quality factor below array size of 19. This might be because there are two major mechanisms that affect the quality factor. When the array is infinitely extended, the quality factor is expected to reach its maximum value. By decreasing the array size, the quality factor continuously decreases due to boundary effect. However, in this regime, the mode profile remains minimally perturbed. On the other hand, below as certain size, the mode profile becomes

$i = 1, 2, 3$ indicates the coupling of any site at the position of $r$ that belongs to the sublattice $A$ to its three nearest neighbors from sublattice B [25]. The operators $a_r$ and $b_r$ act on sublattices A and $B$, respectively. The coupling detuning parameter is defined as

$$\delta\kappa_{r,i} = \Delta(r)e^{jK_+\cdot i}e^{jG\cdot r}/3 + c.c., \quad (2)$$

with $K_\pm = \pm(4\pi/3\sqrt{3}p, 0)$ being the position of the corners in the reciprocal space where the Dirac points appear. Finally, the wave vector $G = K_+ - K_-$ couples two Dirac points at $K_+$ and $K_-$. To obtain the required coupling detuning, the position of the elements should be displaced according to [25]

$$\Delta(r) = d(r)\begin{Bmatrix} sin(k\cdot r + \phi(r))\hat{x} \pm \\ cos(k\cdot r + \phi(r))\hat{y} \end{Bmatrix}, \quad (3)$$

where $d(r) = d_0 \tanh(|r|/\zeta)$. In the aforementioned equations, the quantities $d_0$ and $\zeta$ represent two design parameters that can control the modulation strength of the Kekulé texture, thus controlling the quality factor and mode size of the cavity. For example, a larger $d_0$ will result in a more confined mode with an enhanced quality factor, while a larger $\zeta$ will result in a larger mode size. Adjusting these parameters will allow one to design various MBS cavities for different applications such as lasers exhibiting lower thresholds or generating higher output powers. On the other hand, the phase part of $\Delta(r)$ does not play a role on the modulation strength.

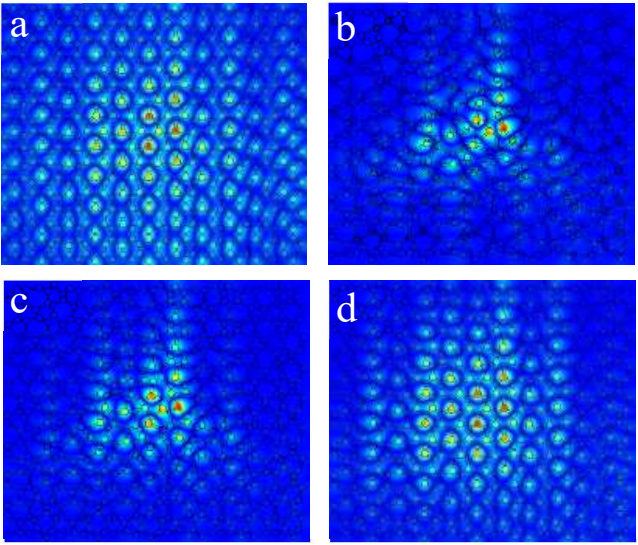

FIG. 4. a,b) Electric field intensity profile of the structure with Kekulé texture when $\zeta = 0.1p$ but $d_0 = 0.05p$ (a), and $d_0 = 0.25p$ (b). By increasing $d_0$ the MBS mode tends to be more confined with a larger quality factor. c,d) Electric field profile of the structure with Kekulé texture when $d_0 = 0.2p$ but $\zeta = 0.1p$ (c), and $\zeta = 6p$ (d). By increasing $\zeta$ the MBS mode becomes more distributed with a smaller quality factor.

distorted too. Therefore, the quality factor is affected by the second mechanism (i.e., mode distortion) and drops rapidly as shown for sizes below 19. For each array size, the minimum required gain for laser is calculated by sweeping the imaginary part of the refractive index of the gain region until the loss of the cavity is fully compensated. Consequently, the required gain threshold shown in Fig. 3d is calculated based the following equation:

$$\gamma_{th} = 2\frac{2\pi}{\lambda} Im\{n_{QW}\} \times 10^{-2} \; [cm^{-1}] \quad (4)$$

In this equation, $n_{QW}$ is the complex refractive index at each resonant wavelength of $\lambda$. Fig. 3d does not account for scattering losses due to etching of the holes. As a result it provides the lower limit for the required gain. It is important to notice that a single-mode cavity is essential for applications such as in lasers and devices that require mode stability in the presence of physical changes that might happen either during fabrication or operation. The Majorana bound state modes can withstand such changes and minimize the sensitivity as long as the vorticity of the texture has not been significantly compromised.

To tune the size of the Majorana bound state and its confinement to the center of the structure, there are two controlling knobs of $d_0$ and $\zeta$. When $d_0$ is small, the structure resembles more of a hexagonal lattice (Fig. 4a), thus the mode tends to remain distributed over the entire array, i.e. the mode size is large because the band opening is small. For an array with a limited extension, this leads to an MBS mode with lower quality factor due to edge termination. Consequently, increasing $d_0$ can enhance the deformation degree of the Kekulé texture (Fig. 4), thus resulting in a more confined mode with larger quality factor at the center of array. On the other hand, $\zeta$ has an opposite effect. By increasing (decreasing) the value of $\zeta$, the deformation degree of the Kekulé texture reduces (increases) as shown in Fig. 4c (Fig. 4d). In all cases, there is a trade-off between the quality factor and the number of cells that are excited by the MBS mode. For a larger quality factor (Fig. 4a and 4d) the number of elements that are involved is less because only central cells have non-zero intensities. However, for a low quality factor, more cells will be excited, thus the cavity can handle higher power levels (Fig. 4ba and 4c). This can be considered when designing a cavity for high power laser purposes that needs to engage more elements with gain materials, or a more coherent laser with a narrow linewidth that needs a larger quality factor.

### III. Perturbation Analysis of MBS

In this section we study the robustness of the MBS cavity to different structural changes using a perturbation theory. For this analysis, the unperturbed Hamiltonian of the structure $\boldsymbol{H}$ which satisfies the eigen-energy relation $\boldsymbol{H}|\boldsymbol{n^{(0)}}\rangle = \boldsymbol{E_n^{(0)}}|\boldsymbol{n^{(0)}}\rangle, \boldsymbol{n} = \boldsymbol{1, 2, 3}, \ldots$ is perturbed with a Hamiltonian $\boldsymbol{H'}$ that represents any imperfection in the system. The perturbation Hamiltonian $\boldsymbol{H'}$ should itself satisfy the Hermitian condition. Thus, the new Hamiltonian has the form of $\boldsymbol{H_p} = \boldsymbol{H} + \boldsymbol{\lambda H'}$, where $\boldsymbol{\lambda}$ is the degree of perturbation that can take any value between 0 and 1.

According to perturbation theory, the eigen-energies of this perturbed system take the form of

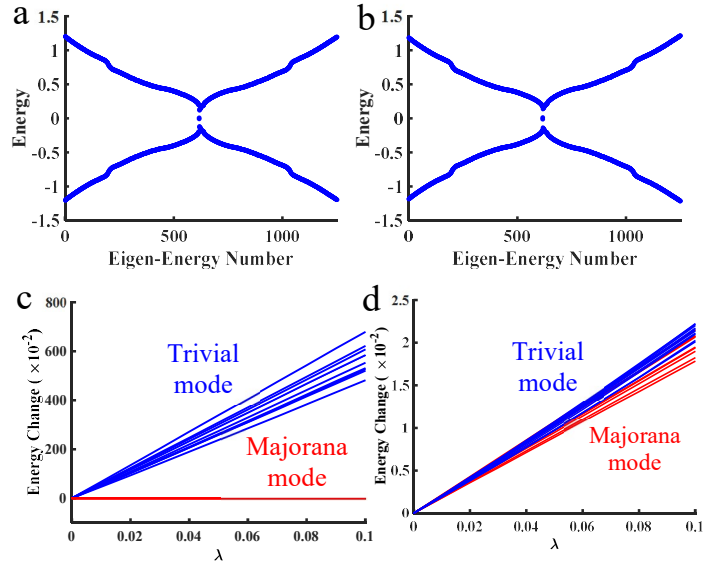

FIG. 5. The robustness of the bandgap and zero mode to perturbation when a) $\lambda = 0.01$ and b) $\lambda = 0.1$. c,d) Variation of the eigen-energy in hexagonal arrangement (blue) and Kekulé texture (red) when there is a perturbation in coupling between cells (c) and changing the sizes of each cell (d). Several lines are from using different random perturbations.

$$E_n = E_n^{(0)} + \lambda E_n^{(1)} + \lambda^2 E_n^{(2)} + \cdots, \qquad (5)$$

where $E_n^{(m)} = \frac{1}{m!}\frac{d^m E_n}{d\lambda^m}|_{\lambda=0}$, and $m = 1, 2, 3, \ldots$.

Here we numerically study the effect of perturbation on the Kekulé Hamiltonian. In our model, the Hamiltonian of $H'$ is generated randomly. For this system, we consider two different sources of perturbations caused by variation in size and inter-element distance of array cells. When the size of the cells is changed, i.e. the on-site resonances associated with the cylinders, the diagonal elements of the Hamiltonian matrix are perturbed. On the other hand, when the distance between the cells varies, it affects the coupling between them, thus changing the off-diagonal elements of the Hamiltonian. In the following, we consider both types of perturbations and their effect on the MBS mode. In addition, we compare the results with trivial modes in the same structures.

Figure 5a and 5b show the band structure with the perturbed Hamiltonian $H_p$ for two different values of $\lambda$, representing weak and strong off-diagonal perturbations. As it is evident in these figures, even by increasing the strength of perturbation, $\lambda$, the bandgap stays open and the zero mode remains locked at the middle of the gap. Figure 5c shows the robustness of the MBS mode when the coupling between the constituent elements of the array is changed as a function of $\lambda$. Compared to the trivial modes shown in Fig. 3b, the zero mode clearly undergoes almost no variation, i.e. it remains stable and robust as perturbation strength ($\lambda$) intensifies. This robustness against variations in exchange interactions (coupling strengths in the lattice or off-diagonal elements of the Hamiltonian) is universally observed in topological systems. However, when the resonance of each cell is changed, both MBS and trivial modes become affected by the perturbation in an almost similar fashion (Fig. 5d). This is also aligned with other topological models like SSH arrays and 2D topological insulator materials in which all modes, trivial and topological, show a similar level of sensitivity to on-site energy variations. In addition, when compared to the trivial modes, the topological Majorana bound state also exhibits larger resilience to local perturbations that affect only a part of the structure.

## IV. Conclusion

In conclusion, we introduced a new type of a 2D topological photonic cavity based on Majorana bound states. The cavity was implemented on a III-V semiconductor platform as a PhC hexagonal lattice, modified by a Kekulé texture. The MBS mode presents the highest quality factor compared to the other modes of the structure and it can scale up by increasing the size of array or changing the Kekulé transformation parameters. These characteristics make this structure a suitable cavity for applications that require more confinements with larger quality factors or higher output powers. In addition, this mode is topologically robust to defects and remains single-moded even in the presence of imperfections that affect the coupling between the array elements. The properties of the mode (size, quality factor, etc.) are controllable by adjusting the Kekulé texture parameters. This type of cavity can be useful in designing novel distributed resonators with better functionalities for lasers and integrated optical circuits where stable and low sensitivy cavities are required.

## References

[1] R. Jackiw and C. Rebbi, Phys. Rev. D 13, 3398 (1976).
[2] R. Jackiw and P. Rossi, Nucl. Phys. B190, 681-691 (1981).
[3] F. Wilczek and A. Zee, Phys. Rev. Lett. 52, 2111 (1984).
[4] Y. Ronen, Y. Cohen, J. H. Kang, A. Haim, M. T. Rieder, M. Heiblum, D. Mahalu and H. Shtrikman, PNAS 113, 1743-1748 (2016).
[5] N. Read and D. Green, Phys. Rev. B 61, 10267 (2000).
[6] D. A. Ivanov, Phys. Rev. Lett. 86, 268 (2001).
[7] J. D. Sau, R. M. Lutchyn, S. Tewari and S. D. Sarma, Phys. Rev. Lett. 104, 040502 (2010).
[8] Y. Oreg, G. Refael, and F. von Oppen, Phys. Rev. Lett. 105, 177002 (2010).
[9] A. C. Potter and P. A. Lee, Phys. Rev. B 85, 094516 (2012).
[10] V. Mourik, K. Zuo. M. Frolov, S. R. Plissard, E. P. A. M. Bakkers and L. P. Kouwenhoven, Science 336, 1003-1007 (2012).
[11] L. P. Rokhinson, X. Liu and J. K. Furdyna, Nature Physics 8, 795-799 (2012).
[12] S. Nadj-Perge, I. K. Drozdov, J. Li, H. Chen, S. Jeon, J. Seo, A. H. Macdonald, B. A. Bernevig and A. Yazdani, Science 346, 602-607 (2014).
[13] M. H. Freedman, Proc. Natl. Acad. Sci. USA, 95, 98–101 (1998).
[14] A. Y. Kitaev, Annals of Physics 44, 131-136 (2001).
[15] A. Y. Kitaev, Physics-Uspekhi 303, 2-30 (2003).
[16] T. Iadecola, T. Schuster and C. Chamon, Phys. Rev. Lett. 117, 73901 (2016).
[17] J. Noh, T. Schuster, T. Iadecola, S. Huang, M. Wang, K. P. Chen, C. Chamon and M. C. Rechtsman, Nature Physics 16, 989 (2020).
[18] C. W. Chen, N. Lera, R. Chaunsali, D. Torrent, J. V. Alvarez, J. Yang, P. San-Jose and J. Christensen, Adv. Mater. 31, 1904386 (2019).
[19] A. J. Menssen, J. Guan, D. Felce, M. J. Booth and I. A. Walmsley, Phys. Rev. Lett. 125, 117401 (2020).
[20] X. Gao, L. Yang, H. Lin, L. Zhang, J.g Li, F. Bo, Z. Wang and L. Lu, Nature Nanotechnology 15, 1012-1018 (2020).
[21] M. Parto, S. Wittek, H. Hodaei, G. Harari, M. A. Bandres, J. Ren, M. C. Rechtsman, M. Segev, D. N. Christodoulides and M. Khajavikhan, Phys. Rev. Lett. 120, 113901 (2018).
[22] H. Zhao, P. Miao, M. H. Teimourpour, S. Malzard, R. El-Ganainy, H. Schomerus and L. Feng, Nat. Comm. 9, 981 (2018).
[23] A. Blanco-Redondo, I. Andonegui, M. J. Collins, G. Harari, Y. Lumer, M. C. Rechtsman, B. J. Eggleton and M. Segev, Phys. Rev. Lett. 116, 163901 (2016).
[24] H. Deng, X. Chen, N. C. Panoiu and F Ye, Opt. Lett. 41, 4281 (2016).
[25] C. Y. Hou, C. Chamon and C. Mudry, Phys. Rev. Lett. 98, 186809 (2007).
[26] M. A. Bandres, S. Wittek, G. Harari, M. Parto, J. Ren, M. Segev, D. N. Christodoulides and M. Khajavikhan, Science 359, 1231 (2018).
[27] G. Harari, M. A. Bandres, Y. Lumer, M. C. Rechtsman, Y. D. Chong, M. Khajavikhan, D. N. Christodoulides and M. Segev, Science 359, 1230 (2018).
[28] B. Bahari, A. Ndao, F. Vallini, A. E. Amili, Y. Fainman and B. Kanté, Science, 358, 636-940 (2017).
[29] M. Parto, S. Wittek, H. Hodaei, G. Harari, M. A. Bandres, J. Ren, M. C. Rechtsman, M. Segev, D. N. Christodoulides and M. Khajavikhan, Phys. Rev. Lett. 120, 113901 (2018).
[30] Y. Zeng, U. Chattopadhyay, B. Zhu, B. Qiang, J. Li, Y. Jin, L. Li, A. G. Davies, E. H. Linfield, B. Zhang, Y. Chong and Q. J. Wang, Nature 578, 246–250 (2020).

[31] J.-H. Choi, W. E. Hayenga, Y. N. G. Liu, M. Parto, B. Bahari, D. N. Christodoulides and M. Khajavikhan, Nature Communications, (2021).
[32] Y. N. G. Liu, P. S. Jung, M. Parto, D. N. Christodoulides and M. Khajavikhan, Nature Physics (2021).
[33] L. H. Wu and X. Hu, Phys. Rev. Lett. 114, 223901 (2015).
[34] D. L. Bergman, Phys. Rev. B 87, 035422 (2013).